\documentclass[12pt,preprint]{aastex}

\slugcomment{.}

\shorttitle{}
\shortauthors{Lieu, Mittaz \& Hillman}

\begin{document}

\title{Are the WMAP angular magnification
measurements consistent with an inhomogeneous critical
density Universe?}

\author{Richard Lieu\altaffilmark{1} and Jonathan P.D. Mittaz\altaffilmark{1}}

\affil{Department of Physics, University of Alabama at Huntsville,
Huntsville, AL 35899}

\begin{abstract}
The propagation of light through a Universe of (a) isothermal mass spheres
amidst (b) a homogeneous matter component, is considered.
We demonstrate by an analytical
proof  that as long as a small light bundle passes {\it through}
sufficient number of (a) at various
impact parameters -  a criterion of great importance -
its average
convergence will
exactly compensate the divergence
within  (b).
The net effect on the light is statistically the same as if all the matter
in (a) is `fully homogenized'.
When applying the above ideas towards understanding
the angular size of
the primary acoustic peaks of the microwave background, however,
caution is needed.
The reason is that most (by mass) of (a) are in
galaxies - their full mass profiles are not
sampled by passing light - at least the inner 20 kpc regions of
these systems are missed by the majority of rays, while the rest
of the rays would map back to unresolvable but magnified,
randomly located spots to compensate for the loss in angular size.
Therefore, a scanning pair of WMAP beams finds most frequently that
the largest temperature difference occurs when each beam is placed
at diametrically opposite points of the Dyer-Roeder collapsed sections.
This is the {\it mode} magnification, which corresponds to the acoustic
{\it peaks}, and is less than the mean (or the
homogeneous
pre-clumping angular size).
Since space was seen to be Euclidean
without taking the said adjustment into account, the true density of the
Universe should be supercritical.  Our analysis gives
$\Omega_m =$ 0.278 $\pm$ 0.040 and $\Omega_{\Lambda} =$ 0.782 $\pm$ 0.040.
\end{abstract}

\vspace{2mm}

\noindent
{\bf 1. Introduction}

The propagation of light through the inhomogeneous near Universe is
an intriguing phenomeon, especially from the viewpoint of the cosmic
microwave background (CMB), because the subject is sufficiently
unfathomable that a large number of papers appeared
in the literature (triggered by the tautological `flux conservation'
argument of Weinberg 1976).
Some of the earlier works were cited in section 9.2 of the
review of Bartelmann \& Schneider (2001).  The recent
controversy persists over whether, in a critical density Universe, the
convergence of rays by mass concentrations is balanced by
the divergence in between, so that the average behavior of
the light continues to reflect zero curvature space (see e.g.
Holz \& Wald 1998, Claudel 2000, and Rose 2001).

The observational status of the near Universe is that it comprises
a smooth component which
harbors $\sim$ 35 \% (Fukugita 2003; Fukugita, Hogan,
\& Peebles 1998) of the $\Omega_m =$ 0.27
total matter density (Bennett et al 2003), plus mass clumps
with (to zeroth order) a
limited  isothermal sphere density profile $\rho \propto 1/r^2$
for $r \leq$ some cutoff radius $R$, which are
the galaxies, groups, and clusters.  In the present work  we
demonstrate that the problem concerning the
mean convergence of light in a Universe of
isothermal spheres placed within an otherwise homogeneous space
can be solved analytically.

\vspace{2mm}

\noindent
{\bf 2. Cross section evolution of a light bundle from the Sach's
optical equations}

Let us express the normalized matter
density parameter for the near Universe as
$\Omega_m=\Omega_h+\Omega_g$, where $\Omega_h$ represents  the
homogeneous  component and $\Omega_g$ an ensemble of
uniformly but randomly placed
isothermal spheres.

The framework of our treatment is the
Friedmann-Robertson-Walker
(FRW) space-time as shaped by the homogeneous component of
the matter distribution.  Upon this metric we envision a null geodesic
directed along the backward light cone from the spatial
origin at the observer, whose clock keeps the present time.  The standard
solution 
of the geodesic equation (Eq. (14.40), Peebles 1993) gives
\begin{equation}
\dot t=-\frac{a_0}{a}=-(1+z),
\end{equation}
where the dot derivative is w.r.t. the affine parameter $\lambda$, $a(t)$ is
the (Hubble) expansion parameter at world time $t$, and the initial condition
$\dot t=-1$ at $z=0$ was applied along with the notational shorthand $c=1$.

Now consider small excursions of the actual
light path from a given
radial null geodesic.  It is convenient to introduce
transverse coordinates $\vec l=(l_\alpha)_{\alpha=1,2}$,
such that
 \begin{equation}
d\vec l^2=dl_\alpha dl^\alpha=a^2r^2
(d\vartheta^2+\sin^2\vartheta\,d\varphi^2).
 \end{equation}
Take in particular a pair of backward null geodesics
starting from the origin at the present time.  Let their
separation at affine distance $\lambda$ be $\delta\vec
l(\lambda)$.  The rate of change of this separation is
governed indirectly by the Sachs scalar optical equations, which in the
(reasonable) limit of vanishing Weyl tensor read:
 \begin{eqnarray}
\dot\theta&=&-\theta^2-\frac{1}{2}w^2-
\frac{1}{2}R_{\mu\nu}u^\mu u^\nu,\\
\dot w_{\alpha\beta}&=&-2\theta w_{\alpha\beta}.
 \end{eqnarray}
where 
\begin{equation}
u^\mu=\dot x^\mu=(\dot t,0,0,\dot r),
\end{equation}
and
$w^2=w_{\alpha\beta}w^{\alpha\beta}$.  Here $\theta$ is the
expansion and $w_{\alpha\beta}$ is the shear (the latter is a symmetric
and traceless tensor) - they are quantities which determine the
evolution of $\delta\vec l(\lambda)$ via the null Raychaudhuri equation
 \begin{equation}
\delta \dot l_\alpha=\theta\delta l_\alpha
+w_{\alpha\beta}\delta l_\beta.
 \end{equation}

In Eq. (3) the Ricci tensor $R_{\mu \nu}$ is obtained from the Einstein's
field equations with
the stress energy tensor $T_{\mu \nu}$ having $T_{00}
= \rho$ as the only non-vanishing entry.
The solution, $R_{00} = R_{33} = 4 \pi G \rho$, is well known,
and may be coupled with
Eqs. (5) and (1) to yield an expression for the Ricci focussing source term  as
$R_{\mu\nu}u^\mu u^\nu=8\pi G\rho(1+z)^2$.  Hence Eq. (3) may be written as
\begin{equation}
\dot\theta=-\theta^2-\frac{1}{2}w^2-
\frac{3}{2}H_0^2\Omega_h(1+z)^5
\end{equation}
For the present purpose it is only necessary to work with the two scalar
variables $\theta$ and $w^2$, the evolution of the latter is according to
the equation
 \begin{equation}
\frac{d}{d\lambda}(w^2)=-4\theta w^2.
\label{wdot}
 \end{equation}
which is obtainable from Eq. (4).

In the next step, we suppose
that light from a source at affine distance $\lambda_s$ passes
through one
single isothermal sphere of mass
$M$, centered at $\lambda_l$, and with an impact parameter
$b$ (a physical distance measured at the
lensing epoch).
The angle of deflection $\psi(b)$ is given by:
 \begin{equation}
\psi(b)=\frac{4GM}{R}\left[\arccos\left(\frac{b}{R}\right)+
\frac{R-\sqrt{R^2-b^2}}{b}\right]\qquad(b \leq R),
 \end{equation} 
and
 \begin{equation}
\psi(b)=\frac{4GM}{b}\qquad(b>R).
 \end{equation}

From Eq. (6) we see that the changes
in $\theta$ and $w_{\alpha\beta}$ due to the presence of
the lensing mass are of the form:
 \begin{eqnarray}
\delta(\theta+w_{\rho\rho})&=&-(1+z)\frac{d\psi(b)}{db},\\
\delta(\theta+w_{\phi\phi})&=&-(1+z)\frac{\psi(b)}{b},
 \end{eqnarray}
where $z$ is the redshift for the epoch of interaction.
The factor of $(1+z)$ arises because of the relation
between $\delta\lambda$ and the proper distance.  It
follows that
 \begin{equation}
\delta\theta = -\frac{1+z}{2}
\left[\frac{\psi(b)}{b}+\frac{d\psi(b)}{db}\right],
 \end{equation}
Substituting Eqs. (9) and (10) into Eq. (13), we obtain
\begin{equation}
\delta \theta = 0~~{\rm for}~b > R;~~
\delta\theta=-\frac{(1+z)}{2}\frac{4GM}{R}\arccos\left(\frac{b}{R}\right)~~
{\rm for}~
b \leq R.
 \end{equation}
For the shear $w^2$, the calculations are more complicated.  Yet the
quantity is easily shown to assume importance only in the strong lensing limit,
i.e. in the present context we can ignore it.

We have to compute the average effect of
all the mass inhomogeneities.  The number density of the isothermal spheres
(neglecting evolution and assuming uniform distribution in
{\it total} density FRW space) is
 \begin{equation}
n=n_0(1+z)^3,\qquad n_0=\frac{3H_0^2\Omega_g}{8\pi GM}.
 \end{equation}
The probability of finding a clump with center at the
position $(\lambda, b)$ to within small ranges
$d\lambda,db$ is
 \begin{equation}
P(\lambda,b)\,d\lambda\,db
=2\pi n_0(1+z)^4\,d\lambda\,b\,db.
 \end{equation}

Since the expansion is {\it additive}, 
the globally averaged change of $\theta$ with $\lambda$ is, from
Eqs. (14) and (16),
 \begin{equation}
\left\langle\frac{d\theta}{d\lambda}\right\rangle_g
= -\frac{3H_0^2\Omega_g}{4GM}(1+z)^5
\int_{b_{\mathrm {min}}}^R
\frac{2GM}{Rb}\arccos\left(\frac{b}{R}\right)\,b\,db
= -\frac{3}{2}H_0^2\Omega_g(1+z)^5
\nonumber\\
 \end{equation}
where at the last step the integral was evaluated with
$b_{\mathrm {min}} \ll R$ in mind.

Putting together the effects of both smooth and clumped
matter, we find for $\theta$, from Eqs. (7) and (17)
the equation
 \begin{equation}
\langle\dot\theta\rangle=-\theta^2-\frac{1}{2}w^2
-\frac{3}{2}H_0^2\Omega_h(1+z)^5
-\frac{3}{2}H_0^2\Omega_g(1+z)^5.
 \end{equation}
The contribution from the $w^2$ (shear) term may be estimated by noting
that, from Eq. (11) and (12), 
 $$
\delta(w^2) = \frac{(1+z)^2}{2}\left[
\frac{\psi(b)}{b}-\frac{d\psi(b)}{db}\right]^2.
 $$
Hence, using Eqs. (9), (10) and the fact that, because of the
random orientation of $\vec{l}$, $w^2$ is additive, we arrive after
integration w.r.t. $\lambda$ at
\begin{equation}
\langle w^2\rangle \sim H_0^2 \Omega_g \frac{GMx_s}{R^2}.
\end{equation}
where $x_s$ is the Euclidean distance to the source as measured at $z=0$.
When compared with the last term of Eq. (18), however, we see that
the shear from the mass concentrations remains relatively unimportant until
$GMx_s/R^2 \geq$ 1, i.e. violation of
the weak lensing criterion
(also the value of $w^2$ for homogeneous matter is zero).  
As long as the lensing is weak,
then, one may ignore the
2nd term on the right side of Eq. (18).  The outcome is that the
expansion of a light bundle depends only on the
total matter content, and {\it not} on the degree of
homogeneity of space.  If space
is Euclidean the solution of Eq. (18) is
 \begin{equation}
\theta = -(1+z)H(z) + \frac{(1+z)^2}{a_0 r} = \frac{1}{ar} \frac{d(ar)}
{d\lambda}.
 \end{equation}
This is
in full agreement with the expected value of the angular diameter distance
at zero curvature, viz. $a(t) r$.

The chief conclusion of this section may
also be obtained
by first directly integrating each
percentage angular magnification $\eta = \psi (x_s-x)x/[2(1+z)x_s b]$
over $dP = 2\pi n_0 [1+z(x)]^2 bdb dx$, the latter
because of randomly
located lenses in an inhomogeneous critical density Universe.  Here
$x$ and $x_s$ are respectively critical
density FRW physical distances at the present epoch (the same meaning as
$a(t)r$ in
the line element of Eq. (2) with $t=t_0$), to a lens
and the source.  We find
\begin{equation}
\langle\eta\rangle = \int \eta dP = \frac{3}{2} \Omega_g H_0^2
\int_0^{x_f}dx\,
[1+z(x)]\frac{(x_s-x)x}{x_s},
\end{equation}
where $x_f$ is the distance to the furthest lens, beyond which space
is smooth.
Then, it has been shown
(Lieu \& Mittaz 2005) that, irrespective of the lensing strength,
$\langle\eta\rangle$ {\it is exactly equal to the
demagnification due to the Dyer-Roeder (DR)
beam divergence} in between these encounters, Dyer \& Roeder
(1972).  This method of proof, though no less valid,
is not as elegant in that the two
counteracting effects have to be calculated separately, and
subtracted from each other afterwards.

\vspace{2mm}

\noindent
{\bf 3. Interpretation
of the results, flux conservation; comparison with observations}

We must now understand what the
result of section 2 means.  In particular, we need to know when
the integration over a cylindrical probability
element like that of Eq. (16) corresponds to observational reality.
Among the isothermal spheres of
different scales, viz. galaxies, groups,
and clusters, the first encompasses by far the lion's share of the matter
budget at low $z$,  with $\approx$
50 \% of the entire matter content clumped into
this kind of  large scale structures, i.e.
\begin{equation}
\Omega_g \approx \frac{\Omega_m}{2}~{\rm for~galaxies}  
\end{equation}
(Fukugita 2003, and Fukugita,
Hogan, \& Peebles 1998).  Thus, in order to secure the precarious balance
between beam convergence and divergence, a light signal must
pass through sufficient numbers of galaxies - sampling
the full range of impact parameters - larger systems like clusters
have too small an associated $\Omega$ to play a significant role.
From the observed density of galaxies 
\begin{equation} 
n_0 = 0.17h^{-3}~{\rm Mpc}^{-3} = 0.06
~{\rm Mpc}^{-3}~{\rm for}~h=0.71 
\end{equation}
(Ramella et al 1999) one estimates that throughout the 3 Gpc distance between
$z=0$ and $z=1$ a typical light ray is within $\approx$ 40 kpc from only
one galaxy.  If these 
isothermal spheres cutoff at $R \approx$ 20 kpc (which implies a circular
velocity $\sim$ 250 km s$^{-1}$ using the value of $M$ deduced from
Eqs. (22) and (23)), then
for a separation $\Delta \geq$ 40 kpc
the only `clump' contribution to the
beam expansion will
come from the shear term $w^2$ of Eq. (19) with
$R$ replaced by $\Delta \approx$ 40 kpc.  This calls for
an insignificant correction to $\dot{\theta}$.

The conclusion is that despite the euphoria arising from
Eqs. (18) and (20), most
light rays experience to lowest order only the gravity of homogeneous matter.
What are the ramifications?
Specifically what will the appearance of
features be on a large scale?
Consider  a sequence of small and
contiguous emission pixels on the outlining
contour of an emitting source.  If the
ray bundles connecting them and the observer miss the clumps, their expansion
will evolve according to Eq. (18) without the last term and with a
negligible second term.  This is precisely the
`partially loaded' DR
beam.  It implies demagnification of the pixels
in question.  By Liouville theorem, the pixels remain adjoint, so
to prevent the entire segment
from shrinking they must be
tangentially sheared and pushed back
outwards by the clumps within.  In other words,
when the bulk of a {\it randomly}
located source
boundary is shrunk, it can be restored to original
shape only if the enclosed foreground matter acts as a systematic gravitational
lens.  Yet this is clearly an absurd scenario.
In fact, given that the
clumps are uniformly distributed on either side of any
boundary ray, there is no preferential deflection of the ray,
i.e. concerning most of the boundary which is demagnified
by the DR beam,  the pixels involved should not on average
be mobilized radially inwards or outwards by shear  - this is
consistent with the smallness of the $w^2$ term.
The situation is quite unlike a more homogeneous Universe where
each ray passes through enough representative matter and 
(from section 2) all pixels are
magnified, thereby enlarging the boundary without any center of
radial migration.

If under the `Poisson
regime' of clump distribution
large sections of 
the main boundary of an extended source shorten without distortion
how may this be reconciled with the expected
source flux?  Since
lensing conserves surface brightness, a smaller source means
less detected flux, yet from section 2 we saw that
on balance the effects of lensing and the DR beam cancel, i.e. the flux
(or source size) should be unchanged by clumping.
The answer comes from that minor fraction of the boundary rays which
{\it do} go through clumps.  From the figures given in and after Eq. (22),
we found that this is $\approx$ 25 \%.  The segments involved here
are substantially
enlarged, leading to bulges on randomly located
portions of the boundary, such that
the perimeter now acquires sufficient total length
to enclose a re-magnified area.

We apply the above development to the CMB observations, which
have conventionally been modeled in terms of a critical
density FRW Universe, even though
during the `last leg' of
the light propagation, the matter at low redshift is anything but smooth.
In a more accurate picture, we assume that within $z=1$ some of the matter
is clumped into galaxies, which
have properties as given in Eqs. (15), (22) and (23),
since galaxies
exhibit no evidence for significant evolution up to this redshift (Ofek,
Rix, \& Maoz 2003).
At earlier epochs, the effect of  mass clumping is completely ignored, i.e.
the Universe is treated as homogeneous, and
the possibility of masses missed by the
propagating light above $z=1$  will not be taken into account.
This
understates the outcome of our analysis, which is:
the angular scale of temperature variation
like the CMB primary acoustic peaks (hereafter
PAPs in short) must in the circumstance
demagnify by the percentage expected from a `half-loaded' DR beam
between $z=0$ and $z=1$.   From the reasoning at the end of section 2, we
see that the required quantity is $\langle\eta\rangle$  of Eq. (21)
with $\Omega_g = \Omega_m/2$, $x_f =$ 3.3 Gpc ($z_f =$ 1), and $x_s =$
14.02 Gpc.  The percentage of shrinking
is then $\langle\eta\rangle =$ 10 \%.

When a pair of WMAP TT cross correlation beams
surveys the CMB sky
to measure temperature differences at some beam separation, it most
frequently finds that the largest temperature difference occurs when
each beam is placed at diametrically opposite points of the DR collapsed
contour sections.  This is the {\it mode} magnification, and corresponds
directly to the acoustic {\it peaks}.  Some of the time, however, the
maximum temperature difference is seen at larger angular separations, when 
e.g. one beam is on a point of the DR demagnified section while the other
is on a bulge (or lensed section).  This makes the distribution skewed.
As a result, the mean magnification remains at the pre-clumping value.
Yet the mean is not relevant, for it is the peak position that determines the
total density of the Universe.  Although each beam cannot resolve the lensed
and unlensed sections of the contour (e.g. the former is $\approx$ 0.1 arcmin
in size for a galaxy lens of $R \approx$ 20 kpc, while the beam width is
$\sim$ 30 arcmin for WMAP), that does not change the conclusion.
All it means is that the skewed distribution becomes blurred after convolution
with the beam, the mode stays at its DR demagnified position.

We determined, as is illustrated in Figure 1,
the CMB parameters
required to match the data from WMAP's TT cross correlation power
spectrum, after including the effect of a 10\% 
systematic shift (towards smaller
sizes relative to the pre-clumping value) in the spherical harmonic
number of all structures within the harmonic range of the PAPs.  Since
the observed angular size is Euclidean, one now expects the 
best fit total density to
be supercritical.  They are found at
$\Omega_m =$ 0.278 $\pm$ 0.040 and $\Omega_{\Lambda} =$ 0.782
$\pm$ 0.040, i.e.
$\Omega =$ 1.06.
Given our estimate of the bias by which the PAP
positions represent the true mean density of the Universe,
we also fitted the WMAP data with a smaller bias,
$\langle\eta\rangle =$ 5 \%.  The
figures so obtained are 
$\Omega_m =$ 0.275 $\pm$ 0.040 and 
$\Omega_{\Lambda} =$ 0.755 $\pm$ 0.040, i.e.
$\Omega =$ 1.03.  The other, perhaps even more interesting,
point is that the skewness of the angular size distribution induced
by galaxy lensing, which separates the peak from the mean at this
$\approx$ 10 \% level, is not apparent in the WMAP data, because the
PAPs are symmetric gaussians.

\begin{figure}
\centerline{\includegraphics[height=3.0in,angle=0]{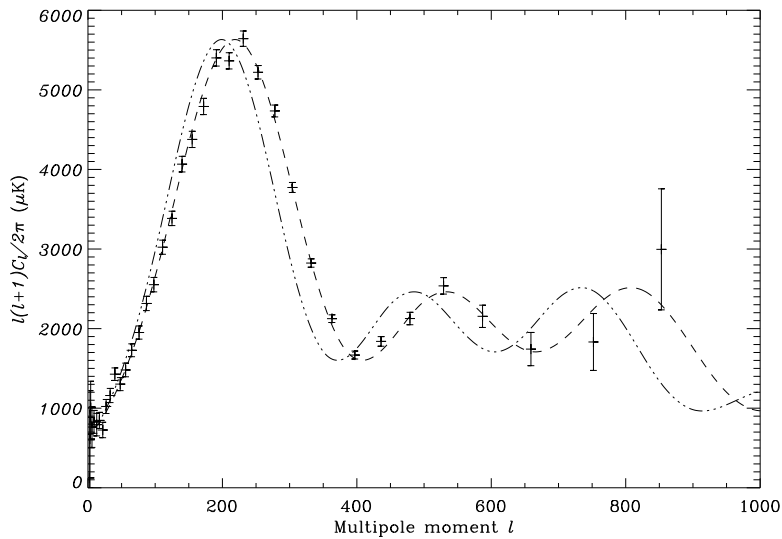}}
\caption{Re-interpreting the
WMAP TT power spectrum, taking account of the fact that
the angular size of large structures as determined with
the cross correlation beam are underestimated 
by 10 \% w.r.t. the homogeneous (pre-clumping) benchmark.
Our approach involved using the CMBFAST code to generate a
model spectrum,
then shifting it
by 10 \% to the right (i.e. towards higher values of $\l$, or
smaller size) and adjusting the
parameters so that the match with the data is secured.
The new
parameters are $\Omega_m =$ 0.278 $\pm$ 0.040, 
$\Omega_{\Lambda} =$ 0.782 $\pm$ 0.040, spectral
slope = 0.975 $\pm$ 0.030, and Hubble constant $h =$ 0.72 $\pm$ 0.03.
Goodness of fit is
$\chi^2 =$ 41.7 for 38 degrees of freedom.  This model, when used
to make predictions for a homogeneous Universe, is shown on the plot
as the dot-dashed curve.  When shifted by 10 \% to account for the effect of
inhomogeneities as discussed above, it becomes the dashed curve.}
\end{figure}

Authors are indebted to Tom Kibble
for helpful discussions.

\vspace{2mm}

\noindent
{\bf References}

\noindent
Bartelmann, M. \& Schneider, P., 2001, Physics Reports, 340, 291.

\noindent
Bennett, C.L. et al, 2003, ApJS, 148, 1-27.

\noindent
Claudel, C. -M., 2000, Proc. Roy. Soc. Lond., A456, 1455.

\noindent
Dyer, C.C., \& Roeder R.C., 1972, ApJ, 174, L115.

\noindent
Fukugita, M., 2003, in `Dark matter in galaxies', IAU Symp. 220, Sydney
(astro-ph/0312517).

\noindent
Fukugita, M., Hogan, C.J., \& Peebles, P.J.E., 1998, ApJ, 503, 518.

\noindent
Holz, D.E., \& Wald, R.M., 1998, Phys. Rev. D, 58, 3501.

\noindent
Lieu, R., \& Mittaz, J.P.D., 2005, ApJ in press (astro-ph/0308305).

\noindent 
Ofek, E.O, Rix, H. -W., \& Maoz, D., 2003, MNRAS, 343, 639.

\noindent
Peebles, P.J.E., 1993, {\it Principles of Physical Cosmology}, Princeton
Univ. Press.

\noindent
Ramella, M. et al 1999, A \& A, 342, 1.

\noindent
Rose, H.G., 2001, ApJ, 560, L15.

\noindent
Weinberg, S., 1976, ApJ, 208, L1.

\vspace{2mm}

\noindent
{\bf Note added in Proof} (though too late to appear in the
ApJL article itself):  Lyman Page was among several who questioned whether
the galaxy lensing bias effect we discussed could still lead to
an acceptable match between the $\Omega =$ 1 standard model and
the WMAP data, if we are prepared to adjust the value of the Hubble constant
$H_0$.  The answer is no.  In fact, such an undertaking yielded a
minimum $\chi^2$ of 401.8 for 28 degrees of freedom - a completely
unacceptable fit.  The best fit values of $H_0$ and $\sigma_8$ then become
72.6 and 0.845 respectively.
discussed 

\end{document}